\documentclass[journal=jpclcd,manuscript=letter,layout=twocolumn]{achemso}

\usepackage[version=3]{mhchem} % Formula subscripts using \ce{}
\usepackage{textgreek}

\author{Anagha Sasikumar}
\affiliation{CIRIMAT, Universit\'e de Toulouse, CNRS, Universit\'e Toulouse 3 - Paul Sabatier, 118 Route de Narbonne, 31062 Toulouse cedex 9, France}
\altaffiliation{R\'eseau sur le Stockage \'Electrochimique de l'\'Energie (RS2E), F\'ed\'eration de Recherche CNRS 3459, HUB de l'\'Energie, Rue Baudelocque,  80039 Amiens, France}
\author{John M. Griffin}
\affiliation{Department of Chemistry, Lancaster University, Lancaster, LA1 4YB, UK}
\author{C\'eline Merlet}
\affiliation{CIRIMAT, Universit\'e de Toulouse, CNRS, Universit\'e Toulouse 3 - Paul Sabatier, 118 Route de Narbonne, 31062 Toulouse cedex 9, France}
\altaffiliation{R\'eseau sur le Stockage \'Electrochimique de l'\'Energie (RS2E), F\'ed\'eration de Recherche CNRS 3459, HUB de l'\'Energie, Rue Baudelocque,  80039 Amiens, France}
\email{celine.merlet@univ-tlse3.fr}

\title{Understanding the chemical shifts of aqueous electrolyte species adsorbed in carbon nanopores}

\begin{document}

\begin{tocentry}
\includegraphics[scale=1.3]{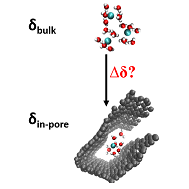}
\end{tocentry}

\begin{abstract}

Interfaces between aqueous electrolytes and nanoporous carbons are involved in a number of technological applications such as energy storage and capacitive deionization. The disordered nature of the carbon materials makes it challenging to characterize ion adsorption and relationships between materials properties and performance. Nuclear magnetic spectroscopy can be very helpful in that respect thanks to its nuclei specificity and ability to distinguish between ions in the bulk and in pores. Nevertheless, several factors can affect the measured chemical shifts making it difficult to interpret experimental results. We use complementary methods, namely density functional theory calculations, molecular dynamics simulations and a mesoscopic model, to investigate various factors affecting the chemical shifts of aqueous electrolyte species. We show that the relative importance of these factors depend on the ion nature. In particular,  the hydration shell has a much more pronounced effect for large polarizable ions such as Rb$^+$ and Cs$^+$ compared to Li$^+$.

\end{abstract}

%\section{Introduction}
% Sections not written in letter

Aqueous solutions under confinement are present and essential in a wide range of chemical and biological processes. Confinement of aqueous electrolytes in carbon nanopores is of particular interest due to its relevance in several technological areas such as elec\-tro\-che\-mi\-cal double layer capacitors (EDLCs)\cite{Fic12,Sajjad21} and capacitive deionization.\cite{Porada13,Luciano20} Confinement alters significantly the properties of the adsorbed species with respect to their bulk counterparts which can in turn affect the performance of these devices. Characterization of quantities such as the hydration structure of the ions in the pores, ion dynamics, ion distribution across pores of different sizes, and ion-carbon wall interactions is thus necessary in order to improve these technologies and has been the subject of numerous studies. 

Experimental attempts to understand the interfacial behaviour of aqueous electrolytes under confinement have involved techniques such as X-ray diffraction,\cite{Ohba12} extended X-ray absorption fine structure,\cite{Ohkubo02} small-angle X-ray scattering,\cite{Prehal15,Prehal17} electrochemical quartz crystal microbalance,\cite{Escobar-Teran22,Wu18} Raman spectroscopy,\cite{Nunes20} nuclear magnetic resonance spectroscopy,\cite{Cervini19,Luo15a,Luo15b} and have provided valuable information about the electrolyte confinement in materials such as carbon nanotubes, slit-shaped nanospaces, graphene nanowrinkles, and nanoporous carbons. However, a quantitative microscopic characterization is still out of reach with many investigation techniques, especially for disordered porous carbons, due to their structural complexity and the existence of fast motion in the systems.

Thanks to its quantitative and nucleus specific features, nuclear magnetic resonance (NMR) spectroscopy is a remarkable technique to study a wide range of systems. The chemical shifts of NMR active nuclei are sensitive to their local environment making this technique compatible with the investigation of aqueous electrolyte species in the bulk and under confinement. More precisely, the peaks observed for species adsorbed to the carbon surface are shifted in the NMR spectrum relative to the species in the bulk electrolyte allowing for a clear identification and quantification of the confined species.\cite{Harris96,Dickinson2000,Griffin14,Griffin16} The additional possibility to conduct NMR measurements while applying a potential difference between the electrodes has also been exploited to study charging mechanisms in EDLCs\cite{Forse16,Griffin14} and even characterize ion diffusion for different states of charge.\cite{Forse17} 

The chemical shift of the species confined in porous carbonaceous materials is mainly the consequence of the secondary magnetic shielding generated by ring currents arising on the carbon surface when applying a primary magnetic field to the system.\cite{Lazzeretti2000,Anderson10} This phenomenon can be gauged using a Nucleus Independent Chemical Shift (NICS) approach.\cite{Chen05,Gershoni-Poranne21} NICS can be evaluated using density functional theory (DFT) calculations and previous studies have shown, amongst other effects, that the shielding effect is larger when the ions are closer to the surface of the carbon.\cite{Xing14,Forse14} Hence the distribution of the ions at the interface and the porosity of the carbon are expected to have a major influence on the chemical shift of the adsorbed species. The local structure, and especially the size of the aromatic domains, was also shown to have a large effect which could be used to get insights into the structures of a range of porous carbons.\cite{Forse15b} 

While the ring currents are predominant in determining the chemical shifts for a range of electrolytes,~\cite{Forse21,Sasikumar21} other parameters can lead to significant modifications of the chemical shifts. NMR spectra are known to be influenced by the hydration number of the ions under consideration, the dehydration of ions upon entering the micropores can thus lead to large chemical shift changes.\cite{Luo15b,Gerken02} Due to exchange between the environments inside and outside of the pores and motional averaging, the carbon particle size can also lead to large variations.\cite{Cervini19} The concentration of ions in organic and aqueous electrolytes adsorbed on microporous carbon can further influence the overall chemical shift.\cite{Cervini19,Fulik18} The interpretation of NMR spectra is, as a consequence, both rich and complex, and distinguishing the contributions from the above mentioned factors is an essential task for a better understanding of the behaviour of the electrolyte species at a local scale. 

\begin{figure}[ht!]
\centering
\includegraphics[scale=0.58]{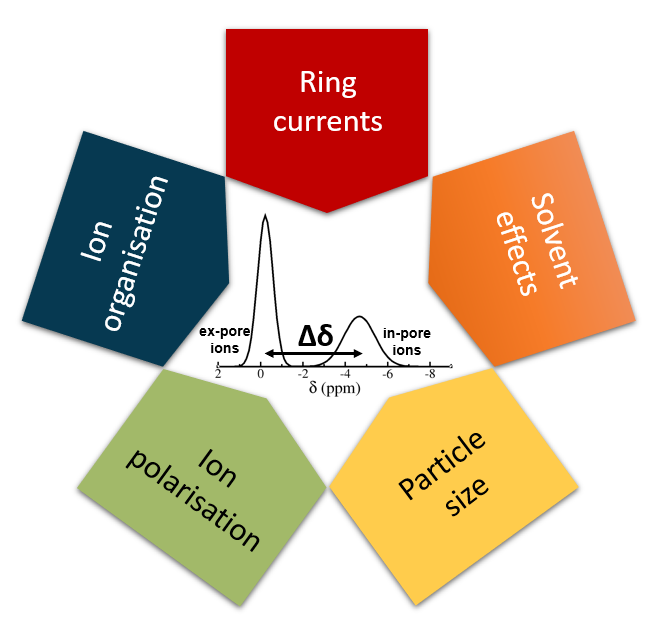}
\caption{Depending on the carbonaceous material and electrolyte, several factors can contribute differently to the chemical shifts of electrolyte species and consequently to the chemical shift difference, $\Delta\delta$, between bulk (ex-pore) and confined (in-pore) ions.}
\label{factors}
\end{figure}

While DFT calculations are adapted to estimate NMR parameters of small static systems, they are not adequate to represent the variety of environments existing in real systems and the ion motion. Molecular dynamics (MD) simulations are more suited for such studies. Feng~\textit{et~al.}\cite{Feng10} conducted si\-mu\-la\-tions of K$^+$ ions in water, in contact with slit pores, and demonstrated that the distribution of these ions in electrified pores of different sizes depend on the ion hydration, the water-water interactions and the pore size. Beckstein~\textit{et~al.}\cite{Beckstein04} used molecular simulations of an aqueous electrolyte in contact with different nanopores and channels to evaluate the energy barriers to Na$^+$ ion permeation, related to ion dehydration, through pores with various radii smaller than 1~nm. Another study on MD simulations of an aqueous NaCl electrolyte and carbide derived carbons (of average pore sizes 0.75~nm and 1.0~nm) have revealed that unlike large organic ions dissolved in acetonitrile, the desolvation of Na$^+$ ions is limited in the carbon nanopores.\cite{ganfoud19} Overall, these works show that different ions are more or less susceptible to dehydration and as a consequence, to fully interpret NMR spectra, one has to consider both the NMR parameters for specific ion configurations and the distribution and exchange of ions between these environments.

In previous works, we proposed to combine the results from DFT calculations and MD simulations using a mesoscopic model to determine NMR spectra.\cite{Merlet15,Sasikumar21} This method takes into account information on i)~ion adsorption and organization, ii)~local magnetic shielding, and iii)~pore size distribution. This technique was shown to give results in good agreement with experimental NMR, Raman spectroscopy and pair distribution function analysis for a range of porous carbons.\cite{Forse15b} The mesoscopic model is also suitable for predicting \emph{in situ} NMR spectra for supercapacitors based on organic electrolytes.\cite{Sasikumar21} One of the assets of the model is to be able to modify independently the ion adsorption and local magnetic shielding entries allowing one to assess the relative importance of these two factors. Thanks to this feature, it was possible to demonstrate that, for organic electrolytes, the variations of the magnetic shieldings, as a consequence of the changes in electronic structure with charging (negatively or positively), have a predominant influence on the overall chemical shift compared to ion organization.\cite{Sasikumar21} This result is consistent with experiments conducted on a range of electrolytes showing similar variations of the chemical shift with applied potential.\cite{Forse21,Fulik18,Wang11b} 

In this work, we investigate aqueous electrolytes with various alkali metal ions in contact with porous carbon particles corresponding to polyether ether ketone derived carbons (PDCs). The motivation for this study is to explore additional effects observed experimentally with aqueous electrolytes compared to organic electrolytes (see Figure~\ref{factors}).\cite{Cervini19} One interesting aspect of the PDCs is that their pore size distributions depend on the activation conditions. Here the PDCs will be referred to using burn-off values which denote the percentage of mass lost after the activation step and is thus indicative of the porosity generated. 

%\section{Results and discussion}
% Sections not written in letter

\begin{figure}[ht!]
\centering
\includegraphics[scale=0.3]{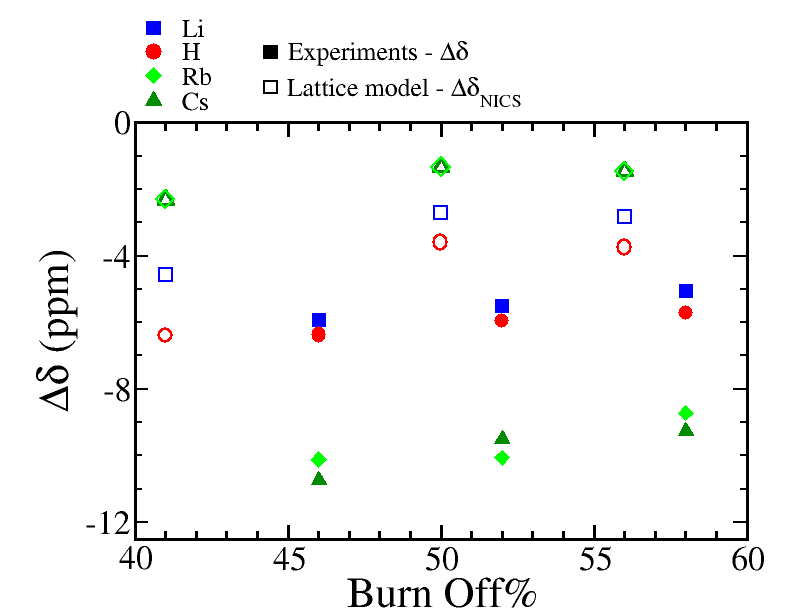}
\caption{Comparison of average chemical shifts of the in-pore alkali metal ions and hydrogen atoms of water relative to the bulk species in various PDCs evaluated by lattice simulations and experiments.\cite{Cervini19} Different symbols and colors correspond to different nuclei probed. Filled symbols correspond to experimental results while open symbols correspond to lattice model results. The $\Delta\delta_{\rm NICS}$ values for Rb$^+$ and Cs$^+$ are superimposed.}
\label{compare_shift}
\end{figure}

The $^7$Li, $^{87}$Rb $^{133}$Cs, and $^1$H NMR spectra for LiCl, RbCl, and CsCl solutions (1M) in various PDCs were simulated using the mesoscopic model previously developed (also referred to as ``lattice model" in the remainder in the article).\cite{Merlet15,Sasikumar21} The information on ion adsorption and organization is provided as free energy profiles (extracted from MD simulations and shown in Figure~S2), the information on local magnetic shielding is given by NICS profiles (calculated through DFT) and the pore size distributions are the ones obtained previously from adsorption isotherms (see Figure~S3).\cite{Cervini19} All data considered correspond to a neutral carbon particle. While the free energy and NICS profiles are calculated in the proximity of a single carbon surface, the lattice model considers slit pores with two carbon surface assuming symmetrical free energies and additive NICS. It is worth noting that any specific interaction or charge transfer between ions / water molecules and the carbon surface are neglected at this stage. The chemical shifts of in-pore species relative to the bulk ($\Delta\delta_{\rm NICS}$) were evaluated and compared against experimental chemical shifts obtained by Cervini \emph{et~al.}.\cite{Cervini19} Results are shown in Figure~\ref{compare_shift}.

In general, the simulated $\Delta\delta_{\rm NICS}$ values decrease with an increase in the burn-off value of the PDC. This is in qualitative agreement with the experiments. The pore size distributions for the PDCs corresponding to the studied range of burn-off values considered show pores in the micropore and meosopore ranges with a large amount of pores around 0.8~nm, 1.1~nm and 2.2~nm.\cite{Cervini19} With the increase in burn-off value, there is usually an increase of pores in the 2-3~nm range. As DFT calculations show that NICS are larger for small pore sizes\cite{Xing14,Forse14}, an increase of the average pore size is expected to lead to a smaller $\Delta\delta$ values. The data for NICS is included in the lattice model, as well as the pores size distributions, which allows for a reproduction of the $\Delta\delta$ variation with burn-off values.

Looking closer at the trend, the $\Delta\delta_{\rm NICS}$ observed for all the species under consideration at a burn-off value of 50\% are slightly smaller (- 5\%) than that of 56\%. In addition to the $\Delta\delta_{\rm NICS}$ values, the lattice model also gives access to the ions or molecules distributions across pores of different sizes. The ion distributions for Li$^+$ and Cs$^+$ in the three PDCs considered are given in Figures~S3 and~S4 (results for Rb$^+$ are very similar to the results for Cs$^+$). For the PDC with 56\% burn off, the fraction of ions occupying pores sizes in the range 1.9-3.2~nm is larger than for the PDC with 56\% burn off leading to a smaller $\Delta\delta_{\rm NICS}$ value for this PDC compared to the 50\% material.

While the trend with burn-off values is qualitatively similar for the various PDCs, the variation of the chemical shifts with the porosity, i.e. the slope of the curve, is slightly overestimated by the simulations. This discrepancy can come from the hypotheses made in the lattice model regarding the NICS calculated through DFT (\emph{e.g.} averages across NICS profiles calculated for different aromatic molecules, positions at which the NICS are calculated close to the carbon surface, ...) which are described in details in previous works.\cite{Merlet15,Sasikumar21} This could also be due to the difference in the carbon structures used for NMR experiments and adsorption studies to determine the pore size distribution. It is worth mentioning that the determination of pore size distributions is subject to some limitations. Different pore size distributions can be obtained for a given porous carbon depending for example on the probe molecule chosen and the model adopted to interpret the adsorption isotherms.~\cite{Dombrowski00,Neimark09} Overall, the agreement between the lattice model and experiments for the trend across burn-off values is satisfying. 

The $\Delta\delta_{\rm NICS}$ calculated using the lattice model for all cations ($^7$Li, $^{87}$Rb, $^{133}$Cs) are similar and smaller than the ones of $^1$H for all porous carbons considered. The small difference of 1-2~ppm between $^1$H, $^7$Li and $^{87}$Rb/$^{133}$Cs can be ascribed to the ion organization at the interface with carbon. Indeed, the free energy profiles extracted from MD simulations show that hydrogen atoms come closer to the carbon surface than the other species, followed by Li$^+$ and by Rb$^+$ and Cs$^+$ (see Figure~S2). As a consequence, the lattice model predicts that the population of Li$^+$ ions (or hydrogen atoms) in small pores is larger than the one of Cs$^+$ (or Rb$^+$) ions leading to a smaller $\Delta\delta_{\rm NICS}$ value for large ions. This is illustrated in Figures~S3 and~S4 giving the distribution of Li$^+$ and Cs$^+$ ions in pores of different sizes. However, one should keep in mind that the lattice model only takes adsorption into account through average free energy profiles obtained at a planar surface and specific energy penalties related to desolvation could change the picture in very small pores. 

While the lattice model satisfactorily predicts the $\Delta\delta$ values of $^1$H and $^7$Li with respect to the experiments\cite{Cervini19}, the values calculated for $^{87}$Rb and $^{133}$Cs are significantly underestimated, by 4-5 ppm. In addition, while the relative variation in $\Delta\delta$ between $^1$H and $^7$Li is reproduced by the lattice model, the relative difference between these two species and $^{87}$Rb/$^{133}$Cs is completely wrong. This shows that the ion organization at the carbon surface and the resulting distribution in the pores, already taken into account in the lattice model, cannot be responsible for this shift. 

In previous works, it was shown that more than ion organization, notable contributions from ring currents on the total shift are to be expected\cite{Sasikumar21}. In the case of interest here, variations in ring currents may arise due to specific interactions and charge transfer between the electronic density of the alkali metal ions and the carbon. These effects are neglected in the NICS calculations used so far but they can be explored with DFT. 

\begin{figure}[ht!]
\centering
\includegraphics[scale=0.3]{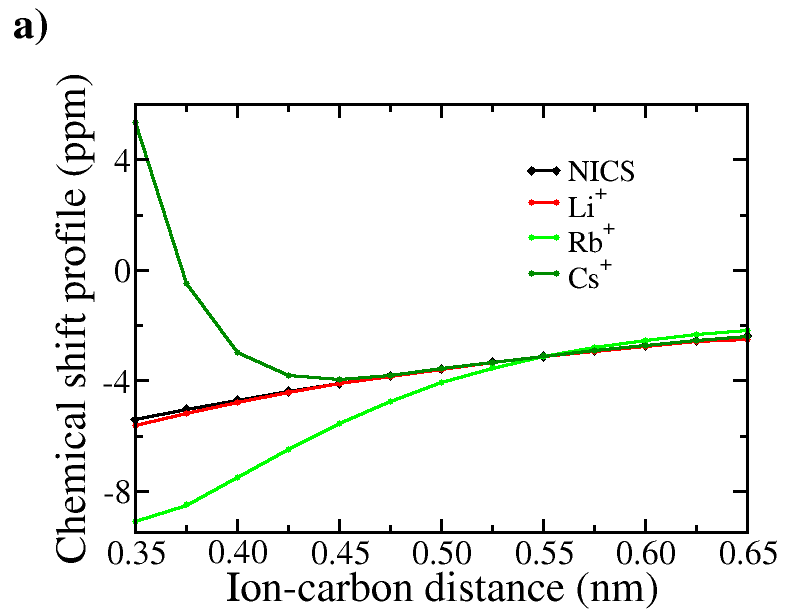}
\includegraphics[scale=0.3]{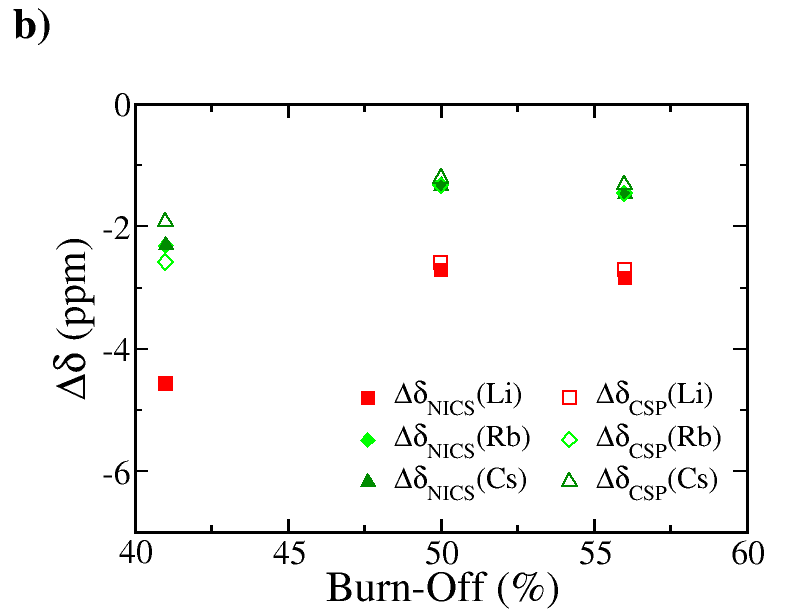}
\caption{a) Chemical shift profiles of various alkali metal ions with respect to vacuum calculated at various distances from the centre of the circumcoronene molecule and NICS values at the same distances. b) $\Delta\delta$ values of the in-pore species relative to the bulk species obtained using the lattice model with the NICS profiles (NICS) and the chemical shift profiles in presence of the ions (CSP).}
\label{shift_M+}
\end{figure}

The chemical shift profiles for Li$^+$, Rb$^+$, and Cs$^+$ ions at various distances from a circumcoronene molecule were calculated by DFT (see Figure~S5). The shifts calculated, referenced to the bare ion in vacuum, are shown in Fi\-gu\-re~\ref{shift_M+}. For completeness, the chemical shifts for other alkali metal ions in the series (Na$^+$ and K$^+$) are given in Figure~S6. For distances below 0.6~nm, the chemical shifts deviate increasingly from the NICS as the ion size increases, which corresponds to an increasing polarizability. The values for Li$^+$ are extremely close to the NICS values. Overall, more negative chemical shifts compared to NICS are observed for the Li$^+$ and Rb$^+$ with the deviation being much larger for the latter. The deviation in the shift is -0.2~ppm and -3.7~ppm for Li$^+$ and Rb$^+$ respectively at a distance of 0.35~nm. The case of Cs$^+$ is strikingly different from the other ions with an increase of shift by 10.7~ppm compared to the NICS at 0.35~nm. 

The origin of the deviation of the chemical shift of the ion from the NICS is not well understood. As the ring currents are sensitive to the molecule charge,\cite{Sasikumar21} a charge transfer between the ion and aromatic molecule could be a reason for such a variation. To test this hypothesis, Bader and Mulliken charges on atoms of the cation - circumcoronene systems were calculated.\cite{g16,bader85,Lu12} The calculations were done for cations 0.35~nm away from the centre of the circumcoronene. The results show evidence of a limited charge transfer of 0.03~e maximum between the cations and the carbon. Interestingly, previous studies have analyzed a small change in the ring currents upon interaction with an ion and the nature of the ion does not affect the ring current changes in the carbon.\cite{Guell05,Rodriguez-otero08,Mary18}. 

One way to evaluate the variation in ring currents due to the proximity of the ion is the calculation of the NICS(1)\cite{Chen05}, i.e the chemical shift of a ``ghost" atom located at 0.1~nm of the ion-circumcoronene system, below the plane of the circumcoronene molecule on the opposite side of the ion. The NICS(1) is evaluated to be -13.6~ppm and -14.0~ppm respectively for Li$^+$ and Cs$^+$. The NICS(1) corresponding to the circumcoronene molecule without any ion nearby is -15.3~ppm. While there is a small change between the systems with and without ion, the variation is probably too small to explain the change in $\Delta\delta$ values for large ions. 

To evaluate the influence on the $\Delta\delta$ values of the variations in chemical shifts observed for large polarizable ions, with respect to NICS, the lattice simulations were repeated with the chemical shift profiles calculated with the actual ions. The results obtained are shown in Figure~\ref{shift_M+}. The $\Delta\delta_{\rm CSP}$ and $\Delta\delta_{\rm NICS}$ values are very similar and as such cannot explain the $\Delta\delta$ values observed experimentally for Rb$^+$ and Cs$^+$. It is worth noting that the free energy profiles for these large ions (Figure~S2) show a first minimum close to 0.6~nm indicating that very few ions get closer to the carbon than this distance. As the chemical shift profiles and NICS values are not significantly different at these distances even for the large ions, it is not surprising that the variations in chemical shifts due to ion-carbon interactions do not lead to noticeable changes in $\Delta\delta$. 

One possibility worth investigating is the fact that MD simulations with planar electrodes, used here to extract free energy profiles, may lead to an underestimation of the amount of ions in smaller pores as shown in other systems.\cite{Sasikumar21} A more realistic representation of the experiment could be obtained from MD simulations with porous carbon electrodes to check the occupancy of the small pores by the cations. This is usually much more involved computationally and less general so it is out of scope of the current work. Instead, MD simulations of the electrolytes confined in slit pores of width 0.8~nm or 1.1~nm were conducted. These simulations do not show any evidence of cations inside the pore with a width of 0.8~nm. In the slit pore of width 1.1~nm, it was observed that Cs$^+$, Rb$^+$ and Li$^+$ ions occupy preferentially the centre of the pore, approximately 0.55~nm away from the carbon surface (see ion and water densities for the in Figure~S9). Interestingly, the MD simulations suggest that the adsorption of Cs$^+$ ions is slightly more favorable than the one of the other cations, in contrast with the lattice simulation results, but the effect is limited. Following these considerations, no major shift is expected from specific ion-carbon interactions. 

Overall, the results from the lattice model, relying on MD simulations and DFT calculations, suggest that ion distributions in pores of different sizes and NICS values are sufficient to explain chemical shifts for Li$^+$ and hydrogen atoms from the water molecules but not for the larger Rb$^+$ and Cs$^+$ ions.

Theoretical calculations have indicated that the hydration number of ions can have a major effect on their chemical shift, this was shown for example for Na$^+$ and F$^-$ ions for various hydration numbers.\cite{Luo15b,Gerken02} This is not accounted for in the lattice simulations. The adsorption and dehydration of ions in hydrophobic carbon pores are dependent on an energy barrier related to the solvation energy of the ions\cite{Luo15b,Beckstein04} and on the pore size. The larger ions, Cs$^+$ and Rb$^+$, are more polarizable than the Li$^+$ ion and have a weaker hydration shell that can distort easily as opposed to the case of Li$^+$. They are therefore more prone to dehydration potentially leading to a significant variation of the chemical shift for confined species. Furthermore, in the vicinity of the carbon, the presence of solvent molecules can affect both the interaction of the ion with the carbon\cite{Rao09} and the carbon itself thus altering the ring currents and the chemical shift. Here, we first investigate the dehydration effect in the bulk before exploring the interface between carbon and ion-solvent complexes. We mainly focus on Cs$^+$ as we expect a similar behaviour for Rb$^+$. 

\begin{figure*}[ht!]
\centering
\includegraphics[scale=0.62]{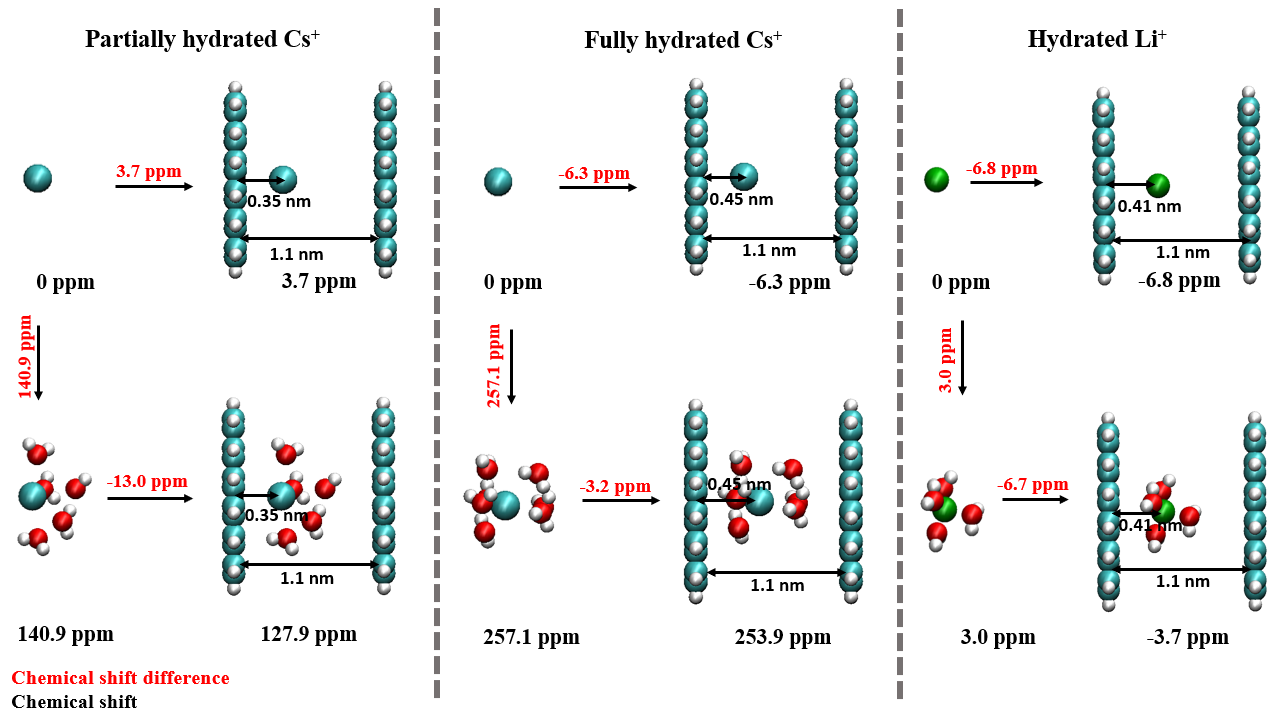}
\caption{Scheme depicting the chemical shift variations for different ion-water-circumcoronene systems corresponding to various local environments for the Cs$^+$ and Li$^+$ ions.}
\label{dehydration_shifts}
\end{figure*}

To study the effect of the hydration number on the chemical shift, four configurations of a fully hydrated Cs$^+$ ion in the bulk were extracted from the MD simulations and the chemical shift of the hydrated ion relative to the bare ion was assessed for different hydration numbers by removing water molecules from the hydration shell (see Supplementary Information for details). Starting with configurations with 9 water molecules (a frequent arrangement in the MD simulations), water molecules were removed one by one until reaching the bare ion state which serves as a reference. Between the full hydration and the bare ion, the variation in chemical shift is huge, reaching almost 200~ppm. It is worth noting that low solvation numbers (below 4) are not present in solution, as can be seen in the MD simulations (see Figure~S10). For hydration numbers between 9 and 6, the variation of the chemical shift with the removal of one water molecule is relatively small ranging from 2 to 10~ppm. For hydration numbers below 6, the difference in chemical shift becomes larger in the range of 10 to 50~ppm. In addition to the effect of the number of water molecules, the calculations indicate that, for a given hydration number, the orientation of the water molecules and the ion-water distances affect the chemical shift significantly. A variation as large as 75~ppm is observed as local structure changes. 

From the DFT calculations, it seems that a dehydration of the ions in the pores could lead to a very large shift. To complement this study, the MD simulations were used to determine the distribution of hydration numbers for the electrolyte species in the bulk and under confinement in a 1.1~nm slit pore. The results are shown in Figure~S10. The distributions for the bulk and in-pore species are superimposed indicating that the extent of dehydration is minimal in this case. The ion densities across the pores for Cs$^+$ and water atoms indicate that the ions mostly reside at the centre of the pore in a fully hydrated state while the water molecules line the pore walls (see Figure~S9). The effect of the water reorientation in the pores can also be neglected as Cs$^+$-O, Cs$^+$-H  pair distribution functions are very similar in the bulk and in the pores (see Figure~S11). It is worth noting that the first coordination shell corresponding to chloride ions is also mostly unaffected by the confinement as can be seen from the Cs$^+$-Cl$^-$ pair distribution functions (see Figure~S11). While the situation could be different in slightly smaller pores, the pore size distributions, showing peaks for 0.8~nm and 1.1~nm, and the fact that ions do not enter pores of 0.8~nm in the simulations suggest that no significant difference in $\Delta\delta$ is to be expected from a variation in the hydration number. 

To go beyond the effect of dehydration alone and include the influence of the hydration in the vicinity of a carbon surface, ion-water complexes for fully and partially hydrated cations were extracted from the MD simulations and DFT calculations of the chemical shifts of the ions next to a circumcoronene molecule were conducted. The distance of the Cs$^+$ ion from the centre of the closest circumcoronene molecule is set to be 0.35~nm for a partially dehydrated state, which was the closest distance between the Cs$^+$ ion and the carbon surface observed in MD simulations, and 0.45~nm for the fully hydrated configuration, which is a distance commonly observed in the MD simulations. To investigate the chemical shift trend in a slit pore, one more circumcoronene molecule is placed parallel to the first such that the pore width is 1.1~nm. For a detailed understanding of the trends, the chemical shifts of bare ions near a carbon wall and in the slit pore were also evaluated. Similar calculations were conducted for Rb$^+$. For Li$^+$, no desolvation is observed in the MD simulations so a single fully hydrated configuration was considered with a circumcoronene - ion distance of 0.41~nm. The most relevant computed shifts are shown in Figure~\ref{dehydration_shifts}. The full set of results is given in Supplementary Information (Figures~S12-S16).

The $\Delta\delta$ value calculated between a partially hydrated Cs$^+$ ion in vacuum, i.e. without the influence of any carbon surface, and the same ion-water complex in a slit pore of 1.1~nm, 0.35~nm away from the closest surface, is -13.0~ppm. This is a larger shift than what is observed in experiments but closer to the expected value compared to a simple NICS or bare ion estimation. Indeed, the $\Delta\delta$ value calculated between a bare Cs$^+$ ion in vacuum and in the same position in a slit pore is 3.7~ppm. For a fully hydrated Cs$^+$ ion in vacuum and at 0.45~nm away from the closest carbon surface, a case observed much more commonly according to MD simulations, the $\Delta\delta$ value calculated is -3.2~ppm. This is this time smaller than the $\Delta\delta$ value for the bare ion of -6.3~ppm. 

Interestingly, the $\Delta\delta$ values calculated for Li$^+$ with or without hydration are very similar (respectively -6.7~ppm and -6.8~ppm). This reinforces the idea that NICS (or chemical shift profiles) are sufficient to estimate the experimental chemical shift difference between bulk and in-pore Li$^+$ ions. On the contrary, for Rb$^+$, large chemical shift differences between hydrated and dehydrated configurations are obtained, in agreement with the fact that NICS (or chemical shift profiles) are, as for Cs$^+$, not sufficient to explain the $\Delta\delta$ values obtained. 

Previous DFT calculations conducted in this work suggest that the change in ring currents is not the main origin for the large shift observed. One way to check this is to use the NICS(1) as was done for the bare cations. The NICS(1) is computed to be -14.3~ppm for the partially hydrated Cs$^+$ ion against -13.9~ppm for the bare ion. Clearly the variation in the chemical shift is not due to the alteration of the ring currents which seems insignificant from this calculation. 

The observations made for various hydration numbers and cations underline the challenge of understanding the factors determining the chemical shifts for large polarizable ions such as Cs$^+$ and Rb$^+$ which are largely affected by their hydration shells. Indeed, experimental results also show a variation of the Cs$^+$ chemical shift with ion concentration even in the bulk,~\cite{Cervini19} an effect we could not explore in this study. To get a better understanding of the relationship between local structure and chemical shift a much broader study including a large number of configurations and, ideally, more realistic nanopores would be needed. Such a study would represent a considerable computational investment and is out of scope of the current work. While the study conducted here on a few configurations is very limited, it proves that $\Delta\delta$ values in the range of -9 to -11~ppm, similar to what is measured experimentally, can be reached when partially or fully hydrated ions go from the bulk to a confined state in a nanopore. 

%\section{Conclusion}
% Sections not written in letter

In this work, we have used a range of computational methods to investigate different factors affecting the chemical shifts of alkali metal ions adsorbed in nanopores with respect to their counterpart in the bulk aqueous electrolytes. Chemical shift differences calculated between the in-pore and bulk ions, the so-called $\Delta\delta$, are systematically compared with experiments conducted with polyether ether ketone derived carbons. A lattice model was first used to explore the importance of the pore size distribution, the ion organization at the carbon surface and the ring currents in the carbon materials in determining the $\Delta\delta$ value. The results show that while this approach seems sufficient to evaluate $\Delta\delta$ for Li$^+$ and hydrogen atoms from water, it leads to very bad estimations for Cs$^+$ and Rb$^+$. DFT calculations realized on ion-water configurations extracted from MD simulations suggest that the hydration shell has a large effect on the chemical shifts for large polarizable alkali metal ions. While the solvation shell of these ions seems mostly unaffected by the confinement, the inclusion of water molecules in the chemical shifts calculations in the vicinity of a carbon surface is essential to reproduce values in good agreement with experiments. In the future, a better understanding of the relationship between local environment and chemical shift would require additional computational studies on a wide variety of ion-water configurations under confinement.

\section*{Methods}
% Sections not written in letter

A thorough interpretation of the experimental NMR spectra of several aqueous electrolytes in PDCs\cite{Cervini19} is attempted here through a combination of classical MD simulations, DFT calculations and a previously developed lattice model\cite{Merlet15,Sasikumar21} which allows one to simulate the NMR spectra of electrolyte species diffusing inside porous carbon materials. In order to model a given system, i.e. a given electrolyte in contact with a PDC, one needs to provide the following information:
\begin{itemize}
\item a pore size distribution;
\item the free energy of ions in pores of different sizes;
\item the chemical shift for ions in pores of different sizes.
\end{itemize}
Following the lattice model calculations, additional MD simulations and DFT were conducted to test possible origins for the discrepancies between calculations and experiments.\\

\emph{Molecular Dynamics simulations}\\

The systems simulated consist in two graphite slabs, placed parallel to each other, either encompassing or surrounded by an aqueous electrolyte in order to represent an unconfined or a nano-confined electrolyte. In the first geometry considered (Figure~S1a), the distance between the graphite slabs is sufficiently large to recover electrolyte bulk properties in the middle of the fluid region. These calculations are used to determine the free energy profiles of ions / water molecules adsorbed at planar carbon surfaces. In the second geometry considered (Figure~S1b), a slit pore, with a pore size of 0.8~nm or 1.1~nm, is immersed in the electrolyte. These calculations are used to investigate confinement effects. The choice of the 0.8~nm and 1.1~nm pore widths is based on the pore size distribution obtained for PDCs through adsorption studies\cite{Cervini19} which show high proportions of pores with these sizes. The electrolytes considered are aqueous solutions of LiCl (1M), RbCl (1M) and CsCl (1M). In all the simulations, 60 ion pairs and 3300 water molecules are used for the electrolyte.

All-atom MD simulations are conducted using the LAMMPS software.\cite{LAMMPS} Intermolecular interactions are represented by the sum of an electrostatic and a Lennard-Jones potential. For the water molecules, the commonly used point charge extended (SPC/E) force field is chosen.\cite{Berendsen87} The parameters for the alkali metal ions, chloride ions and carbon atoms can be found in published works.~\cite{Koneshan98,Cole83} The systems are first simulated in the NPT ensemble at 1~atm and 298~K for 1~ns until they reach a constant volume. Following this NPT step, simulations are conducted in the NVT ensemble at 298~K for at least 20~ns (14.5~ns for RbCl) for the first geometry and 10~ns for the second geometry. The results from the last 19~ns of the first geometry (except for RbCl where the last 12~ns are considered) and last 9~ns of the second geometry are considered for the analysis. A timestep of 1~fs is used for all simulations. Additional details on the simulations are given in Supplementary Information. \\

\emph{Density Functional Theory calculations}\\

All chemical shift calculations are performed using the Gaussian software\cite{frisch2013,g16} and the gauge-including atomic orbital GIAO method. A 6-31G(d) basis set and B3LYP exchange-correlation functional are used for NICS calculations. A 3-21G basis set and B3LYP functional are used for chemical shift calculations involving ions considering the basis set availability and computational expense of the calculations involving heavy atoms such as Cs.  

NICS profiles were evaluated using aromatic molecules (coronene, circumcoronene and dicircumcoronene) as model molecules representing the pore surface. Following previous works, the shielding tensors are calculated at various distances of the carbon surface along a line passing through the centre of the molecule, perpendicular to the molecular plane.\cite{Xing14,Forse14,Kilymis20} The shielding constants are obtained by averaging the diagonal elements of the tensor. The same methodology was used for alkali metal ions at various distances from the centre of the circumcoronene molecule to study the effect of specific ion-carbon interactions (and possible charge transfer) on the chemical shift. The selection of circumcoronene in this case is based on previous simulations which provided a good agreement with the experiments in a range of systems\cite{Merlet15,Sasikumar21}. The results of the calculations in presence of the alkali metal ions are designated as chemical shift profiles.

Chemical shifts of the cations in hydrated clusters, extracted from MD simulations, in the presence and absence of circumcoronene molecules are also determined. To study the effect of the hydration number on the chemical shift, water molecules are removed one at a time starting from a fully hydrated configuration, removing the farthest water molecule first, and the chemical shift is calculated at each step. Several starting configurations were considered to investigate a possible effect of the orientation of neighboring water molecules on the chemical shift of the ions. To evaluate the chemical shifts of fully or partially dehydrated ions in the vicinity of a carbon surface, configurations were extracted from the MD simulations and the chemical shifts were calculated in the presence of circumcoronene at representative distance for the solvation structures considered.\\

\emph{Lattice simulations}\\

In the mesoscopic model,\cite{Merlet15,Sasikumar21} the carbon particles are represented by a cubic lattice, which represents a collection of slit pores. Here, we used 20$\times$20$\times$20 lattice sites. PDCs with burn off values of 41\%, 50\% and 56\% are considered for this study through the inclusion of their pore size distributions obtained experimentally from gas adsorption studies.\cite{Cervini19} Following previous studies, the pore surface distribution is chosen to be a log-normal distribution with a mean of -0.1 and a standard deviation of 0.25 in the case of NICS~\cite{Merlet15,Anouar19,Sasikumar21} and a pore surface corresponding to circumcoronene in the case of chemical shift profiles.

To account for the distribution of ions in pores of different sizes in the carbon material, free energy profiles are obtained from MD simulations. The free energy profiles determined in this work for the species of interest in the range of electrolytes studied are shown in Figure~S2. The shielding environments of the ions adsorbed at each site are defined using the NICS profiles or the chemical shift profiles obtained from DFT calculations. 

The ion and water dynamics between the pores of the carbon particle are modelled through kinetic Monte Carlo moves. The adsorbed species explore different shielding environments (and thus different resonance frequencies) throughout their motion. The NMR signal is calculated for each ion or water molecule, and a Fourier transform gives access to the NMR spectrum and effective chemical shift. A more detailed description of lattice simulations is available in published works\cite{Merlet15,Anouar19}. 

The mesoscopic model allows one to calculate the chemical shift difference, $\Delta\delta$, between species present between particles (bulk) and species present in the pores (in-pore) and to analyse the effects of ring currents and ion organization on the total global shift. It is worth noting that the lattice model does not account explicitly for any effects arising from the hydration of the ions on the global shift.

\begin{acknowledgement} 

This project has received funding from the European Research Council (ERC) under the European Union's Horizon 2020 research and innovation program (grant agreement no. 714581). JMG acknowledges funding from EPSRC with grant number EP/V05001X/1. This work was granted access to the HPC resources of the CALMIP supercomputing centre under the allocations P17037 and P21014, and of TGCC under the allocations A0070911061 and A0080910463 made by GENCI. The authors acknowledge Luca Cervini for sharing experimental data and for fruitful discussions, and Patrice Simon and Alexander Forse for useful conversations.

\end{acknowledgement}

\begin{suppinfo}

Supplementary Information available: details of the molecular dynamics simulations, free energy profiles, pore size distributions and distributions of ions in the different pore sizes, chemical shift profiles for alkali metal ions in the vicinity of circumcoronene, characterization of the ion hydration and its effect on the chemical shift in the bulk, chemical shift for different ion-water-circumcoronene configurations. 

\end{suppinfo}

\section*{Data availability}

The program used to do the lattice simulations is available, along with a manual, on Github (https://github.com/cmerlet/LPC3D). The data corresponding to the plots reported in this paper, as well as input files for the MD and lattice simulations, and DFT calculations, are available in the Zenodo repository with the identifier 00.0000/zenodo.0000000.

\providecommand{\latin}[1]{#1}
\makeatletter
\providecommand{\doi}
  {\begingroup\let\do\@makeother\dospecials
  \catcode`\{=1 \catcode`\}=2 \doi@aux}
\providecommand{\doi@aux}[1]{\endgroup\texttt{#1}}
\makeatother
\providecommand*\mcitethebibliography{\thebibliography}
\csname @ifundefined\endcsname{endmcitethebibliography}
  {\let\endmcitethebibliography\endthebibliography}{}

\end{document}